# Software tool for automatic detection of solar plages in the Coimbra Observatory spectroheliograms

Barata, T. [1], Carvalho, S. [1,2], Dorotovič, I. [3,4], Pinheiro, F. J. G[1]., Garcia, A.[1,5], Fernandes, J.[1,5,6], Lourenço, A. M.[1]

1 Centre for Earth and Space Research of University of Coimbra, 3040–004 Coimbra, Portugal
2 Centre for Mathematics, University of Coimbra, 3001–501, Coimbra, Portugal
3 Slovak Central Observatory, 94701 Hurbanovo, Slovak Republic
4 Computational Intelligence Group of CTS/UNINOVA, 2829–516 Monte de Caparica, Portugal
5 Geophysical and Astronomical Observatory of University of Coimbra, 3040–004 Coimbra, Portugal
6 Department of Mathematics, University of Coimbra, 3001–454, Coimbra, Portugal

**Abstract:**
Full–disk spectroheliograms have been taken in Coimbra on a daily basis since 1926 in the Ca II K–line (K1 and K3). Later, in 1989, with the upgrade of the equipment it was possible to start the observations in the H-alpha line. The spectroheliograms of Coimbra constitutes a huge dataset of solar images, which requires an efficient automatic tool to detect and analyse solar activity features. This work presents a mathematical morphology approach applied to the CaII K3 series. The objective is to create a tool based on the segmentation by watershed transform combined with other morphological operators to detect automatically and analyse chromospheric plages during the solar cycle 24. The tool is validated by comparing its results for cycle 23 with those presented by Dorotovic et al. (2007, 2010). The results obtained are in very good agreement with those, including on images obtained in non-ideal meteorological conditions (eg. some clouds in sky). The results were also qualitatively compared with the results obtained through the application of ASSA model to SDO HMI magnetograms.

**1. INTRODUCTION**

The Sun shows its activity in several ways, like active regions, flares, coronal mass ejections, etc. The characterization of solar features, traditionally made by hand by an expert user, is of great importance to monitor and forecast the solar activity and to obtain results for the Space Weather study (Veronig et al., 2001).

The success of several solar missions has allowed to obtain a vast group of high resolution images. In relation to this available information source, the solar physics community is comparatively small, and therefore the resource to the images digital processing has increased, with the aim of getting information about the solar activity in a prompt and efficient way (Gill et al., 2010; Falconer et al., 2011). Neural networks have been used to detect the solar activity of the solar wind's proton events (Borda et al., 2002), and automatic tracking of solar flares (Caballero and Aranda, 2014). Threshold techniques, region growing, edge detection, segmentation, Hough transform, fractal analysis and fuzzy sets have been applied in the detection of sunspots, active regions, plages, filaments and CMEs (Nesme-Ribes et al., 1996; Benkhalil et al., 2004; Zharkova et al., 2004; Qu et al., 2005; Scholl , 2008; Aboudarham et al., 2008; Fonte and Fernandes, 2009; Gafeira et al., 2013). Hybrid methods that include different approaches have also been developed (Qahwaji and Colak, 2005; Manish et al., 2014; Dorotovic et al., 2014). The mathematical morphology has been applied to sunspots (Curto et al., 2008; Carvalho et al., 2015; Zhao et al, 2016), in the filaments' recognition (Fuller et al., 2005) and plages (Meunier and Delfosse, 2009). A good review is made in by Aschwanden (2010). A common aspect between all these works is the need to incorporate pre–processing techniques, such as Wavelets (Irbah et al., 1999), to normalize the solar images, with regard to dimension, size and intensity, and limb darkening correction (Walton et al., 1998; Denker et al., 1999; Walton and Preminger, 1999; Zharkova et al., 2004; Zharkov et al., 2005).  All these techniques are tailored to detect features in various types of observations at different heights in the solar atmosphere (Verbeeck et al., 2014). Automatic tracking of solar features has been developed for filaments (Gill et al., 2010; Goussies et al., 2010; Higgins et al., 2011) active regions (Pérez–Suárez et al., 2011; Martens et al., 2012), and solar flares (Higgins et al., 2011), coronal mass ejections (Olmedo et al., 2008) which constitute the first steps for the building of an approach that can allow to follow and to characterize the solar activity evolution.

A comparison between automatic and manual methods was made by Zharkova et al. (2005), proving the big efficiency of the automatic methods. Carvalho et al. (2015) compares the results from different sunspot detection methods. The robustness of automatic methods to detect sunspots and active regions are also made by Verbeeck et al. (2013).

Traditionally the solar catalogs were created by hand, but the results of these many applications have contributed for the building of solar activity catalogues, being the EGSO (European Grid of Solar Observations) a good example of this (Fuller et al., 2004). Another pioneer example is the Solar Monitor, which labels active regions using NOAA's (National Oceanic and Atmospheric Association) numbers and heliographic positions (Higgins, 2012).

Despite having more data from new instruments and space missions, it is yet important to maintain older instruments working and to use their data for several important reasons (Hill et al., 2010; Ayres et al., 2012). One of them is the long–term observations of, at least, several decades they have been performing, crucial to understanding the solar cycle. Besides, ground–based observations allow us to preserve and extend consistent data sequences.

Solar faculae or plages are bright areas on the solar surface surrounding active regions and sunspots (Kostik and Khomenko, 2014). They are magnetic structures constituted of flux tubes where a strong magnetic field creates extra heat (about 300 degrees K above surrounding). The interest in facular regions is due the fact they may presage sunspot formation. This interest has increased even more with the discovery that the total solar irradiance increases when the Sun is more active (Solanki and Fligge, 1999; Solanki and Unruh, 2013). In addition, the variability of facular areas it is one of the most important solar indices required to understand the activity of solar cycle (Göker et al., 2016).

This paper intends to contribute towards an automatic detection of facular regions acquired at the Geophysical and Astronomical Observatory of the University of Coimbra during cycle 24. The basic morphological operators are introduced in next section. The data used in this work and the automatic method based on mathematical morphology transforms are described in detail in section 3. Data analysis and discussion of the results are performed in section 4. Finally, the conclusions are presented in the section 5.

## 2. BASIC CONCEPTS OF MATEMATHICAL MORPHOLOGY

Mathematical Morphology is an image analysis theory created in the middle 1960s by George Matheron and Jean Serra in the École des Mines de Paris. Its initial purpose was related to an application in porous media to describe the geometric features of structures (Matheron, 1967). The further developments since then have permitted to construct a solid framework (Matheron, 1975; Serra, 1982) and have successfully reached new application areas (good overview in Soille, 2002), including solar physics (Aschwanden, 2010).

One of the great potentialities of using mathematical morphology is the power to deal with the geometry of complex and irregular shapes (Barata et al., 2015). From the visual analysis of solar images, facular regions present these characteristics which led to exploring an approach based on morphological operators.

Initially developed for binary images this theory was generalized for grey–scale images. Any operator or morphological transform implies the comparison of the features to analyze with a known object, the structuring element. The success of the application of any mathematical transform depends on the choice of the structuring element. The mathematical morphology operators can be used directly or applied sequentially to obtain more elaborated morphologic transformations, for specific ends. Matheron (1975) and Serra (1982) present a detailed description of the mathematical morphology method. In the following paragraphs are presented the main ideas (Soille, 2002).

### 2.1. The basic transforms

The first morphological transforms defined by Matheron (1967), are the erosion ($\varepsilon$) and dilation ($\delta$). To grey scale images, the erosion ($\varepsilon_B(f)$) of an image $f$ by a structuring element $B$ of size $\lambda$, is the minimum of the translations of $f$ by the vectors $-b$ of $B$ (Soille, 2002):

$$\varepsilon_B(f) = \min_{b \in B} f_{-b} \qquad (1)$$

The dilation of an image $(\delta_B(f))$ of an image $f$ by a structuring element $B$ of size $\lambda$, is the maximum of the translations of $f$ by the vectors $-b$ of $B$:

$$\delta_B(f) = \max_{b \in B} f_{-b} \quad (2)$$

The final results for both operators are poor in details when compared with the initial image. If it is necessary to enhance dark areas in a grey image, the erosion shows good results and the dilation has the opposite effect, because white zones in an image are more easily detected.

### 2.2. The morphological gradient

The objective of the morphological gradient is to enhance and extract the contours of the homogeneous regions of grey levels in an image. From the two basic operators, erosion and dilation, the morphological gradient or *Beucher gradient* (Beucher, 1990) is defined as the arithmetic difference between the maximum and the minimum of the function $f$, divided by the size or diameter ($\lambda$) of the structuring element ($B$):

$$\rho_B = \delta_B - \varepsilon_B / 2\lambda \quad (3)$$

The objective of the morphological gradient is to enhance and extract the contours of the homogeneous regions of grey levels in an image.

### 2.3. Opening and closing

The erosion and dilation can be combined to perform two important transforms: opening and closing. The opening ($\gamma$) consists of submitting an image $f$ to an erosion followed by a dilation, both by a structuring element $B$ of size $\lambda$. In the final image the opening cuts peaks and removes small object protuberances:

$$\gamma_B(f) = (\delta_B(f)[\varepsilon_B(f)] \quad (4)$$

In the same manner, the closing transform ($\phi$) consists in applying a dilation to an image $f$ followed by an erosion. This operation suppresses (or closes) all the valleys smaller than a certain dimension given by the size $\lambda$ of the structuring element $B$:

$$\phi_B(f) = (\varepsilon_B(f) \; [\delta_{BB}(f)] \quad (5)$$

Although possible, the isolated application of each operator is not a very powerful filter. However, the alternate application of these operators is the basis of the majority of morphological filters.

### 2.4. Top–hat

The top–hat transform was introduced by Meyer (1979, 1986) based on combinations of openings and closings. The white top–hat (WTH) is defined as:

$$\text{WTH}(f) = f - \gamma(f) \quad (6)$$

and consists in the difference between the original image $f$ and its opening $\gamma$. The white top–hat is used to extract peaks (white regions) in an image with respect to the background. The choice of the shape and the size of the structuring element for both top–hats, depends on the morphology of the structures to be extracted (Soille, 2002).

The black or valley top–hat (BTH), the dual transformation of the WHT, and is the difference of the closing of the original and the original image:

$$\text{BTH}(f) = \varphi(f) - f \quad (7)$$

The closing transform suppress all the valleys smaller than a certain dimension. The subtraction of the initial image allows to recover these structures by putting them at the same level and by simultaneously filtering the ones that were not modified.

**2.5. Geodesic transforms**

Mathematical transforms can be defined either in Euclidean or Geodesic space. Geodesy was introduced in image analysis by Lantuéjoul *et* Beucher (1980). When applied in geodesic space the mathematical transforms are limited to a region or subgroup of the image and only part of the image is analysed (called an image mask). Generally, the geodesic transforms are applied iteratively until stability and this is called reconstruction. This transform is a powerful tool in mathematical morphology and is used to filter images, to find and fill local maxima or minima, to suppress minima and maxima lesser than a give size, to remove noise without affecting structures of interest and filling holes.

The geodesic transforms involve two images: the mask image and the marker image. The marker image is subjected to successive dilations (or erosions) until it fits the mask image (Soille, 2002). The reconstruction by dilation ($R_g$) of a mask image *g* from a marker image *f* is defined as the geodesic dilation of with respect to *g*, iterated until stability (Soille, 2002):

$$R_g^{\delta}(f) = \delta^i(f) \qquad (8)$$

The reconstruction by erosion is the mathematical transform used to eliminate and fill holes in an image. In a binary image, a hole is defined as the set of background components which are not connect to the image border (Soille, 2002) . In grey images the concept of a hole is different and removes holes means eliminate all the minima which are not connected with the image border. To fill holes in a grey image is necessary to impose the set of minima which are connected to the border image. The image marker used in this transform is the set of maximal values on the original image, except along its border where the values of the initial image are kept (Soille, 2002).

## 3. AUTOMATIC DETECTION OF PLAGES OR FACULAR REGIONS
### 3.1. The spectroheliograms

The Astronomical Observatory of the Coimbra University has a collection of solar observations on a daily basis that spans near nine decades until today. Regular observations of the full solar disk in the spectral line of CaII K started in 1926 and in 1989 started the observations in the spectral line Hα. The spectroheliograms from1926 until 2007 were taken through photographical method in the wavelengths 393.37nm (CaII K3), 393.23nm (CaII K1) and 655.87nm (Hα) after 1989. Photographic plates and films 13x18cm were replaced by a 12–bit CCD camera in 2007, making possible to add observations in 656.28 nm (H–halpha continuum) and, after an upgrade of the data–processing software in 2009, to obtain Hα Dopplergrams (Garcia et al., 2010). This extensive collection acquired with the same instrumentation is now available entirely in digital format.

This study is based on CaII K3 series to detect automatically and analyse the chromospheric facular region or plages during the solar cycle 24, from 2008 until 2016.These spectroheliograms are digital images of 8 bits, with 1200 x 1000 pixels. Some examples are shown in Fig. 1. The results obtained were validated by comparing them with results of the study presented in Dorotovič et al. (2007, 2010) for the solar cycle 23. Another evaluation was made by comparing the results obtained by the Automatic Solar Synoptic Analyzer software (ASSA) in SDO HMI magnetograms. The method was also applied in CaII K spectroheliograms of the Astronomical Observatory of Kharkiv State University, with the aim to evaluate the performance of the automatic detection.

### 3.2. Pre–processing the data

Facular regions are bright areas in the solar chromosphere, due the activity of magnetic field. An example of facular regions can be observed on the CaII K3 spectroheliogram acquired at OGAUC in Figure 1. The letters in the corners of the image provide information about the acquisition: data, orientation and place.

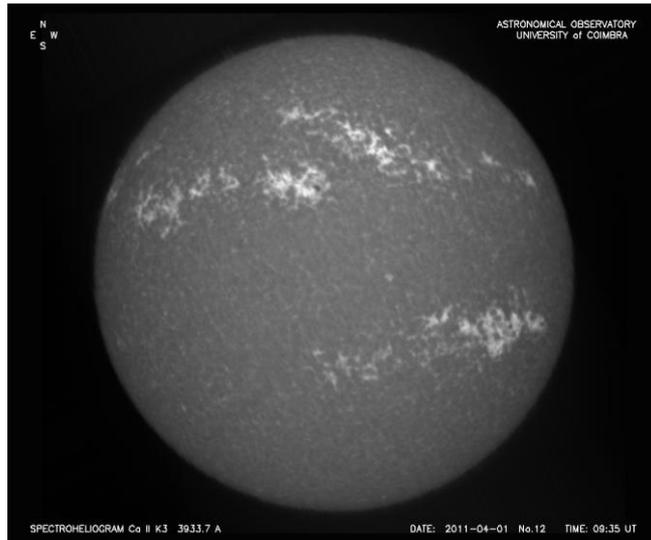

Figure 1 – Example of a CaII K3 spectroheliogram (acquired on the 1st of April 2011).

Prior to the identification of the solar facular regions, it is necessary to define the solar disk within the image. Also, it is necessary to remove the letters because they difficult the application of any automatic processing algorithm. Only the images acquired in digital format (since 2007) have letters in the same position, and the spectroheliograms result from multiple scans of the Sun. Consequently, the edge of the solar disk is an indented surface and the background outside the solar disk heterogeneous. This implies that the digital levels of the background pixels, outside the solar disk, don't have the same value. There is a possibility to have pixels with the same digital level outside and inside of solar disk, which can be a problem to next septs of the automatic detection. Additionally, the solar disk is slightly flatted at the poles. Nevertheless, these pixels of difference can be neglected in comparison to the hundreds of pixels of the solar disk. So, the solar disk can be treated as perfectly circular.

Due to the heterogeneity of the background of the spectroheliograms, a morphological filter is applied, starting by an area closing. This closing operation filters structures without altering the shape of those structures, whose surface area is greater than a given threshold, in this case 70 pixels. Figure 2 (b) shows the result of the application of this transform to the original image of Figure 2 (a). Although the background of the image was filtered, being more homogeneous and less nosily, the letters were not removed. The application of an opening by a disk of size 10 removes the letters (Figure 2 (c)) but does not preserve the original digital levels that corresponds to the solar disk. This can be done by performing a reconstruction, using the original image as a marker (Figure 2 (a)) and the mask (Figure 2 (d)) is the difference between the opening image (Figure 2 (c)) and the original image (Figure 2 (a)). The result of the reconstruction is shown in Figure 2 (e). The binarization of this image (Figure 2 (f)) allows recovering the solar disk and the multiplication of this image by the original, enable to obtain the original digital levels. The result of this pre–processing algorithm is shown in Figure 2 (g), with all the digital level of the background outside the solar disk are equal to 0 (black). This image is the starting point for the automatic recognition of facular regions.

It may seem that this algorithm could be replaced by a binarization applied directly to the original image. However, the heterogeneity of the spectroheliograms background makes impossible to apply an automatic threshold to all images.

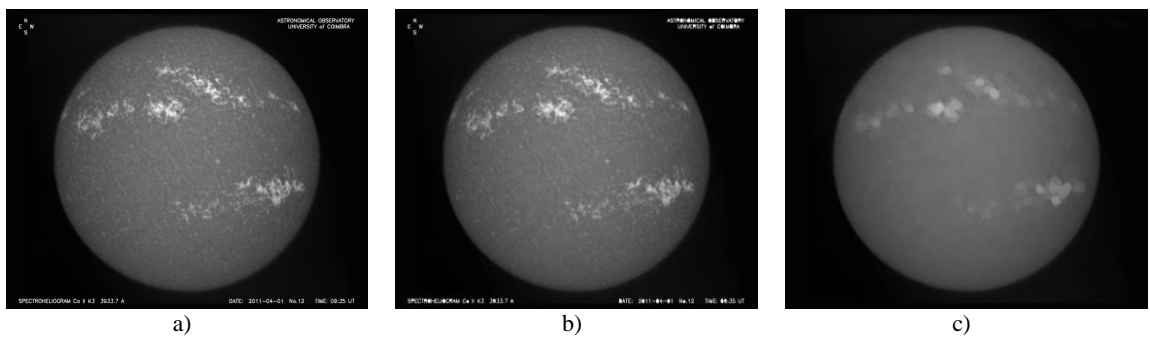

a)                                    b)                                    c)

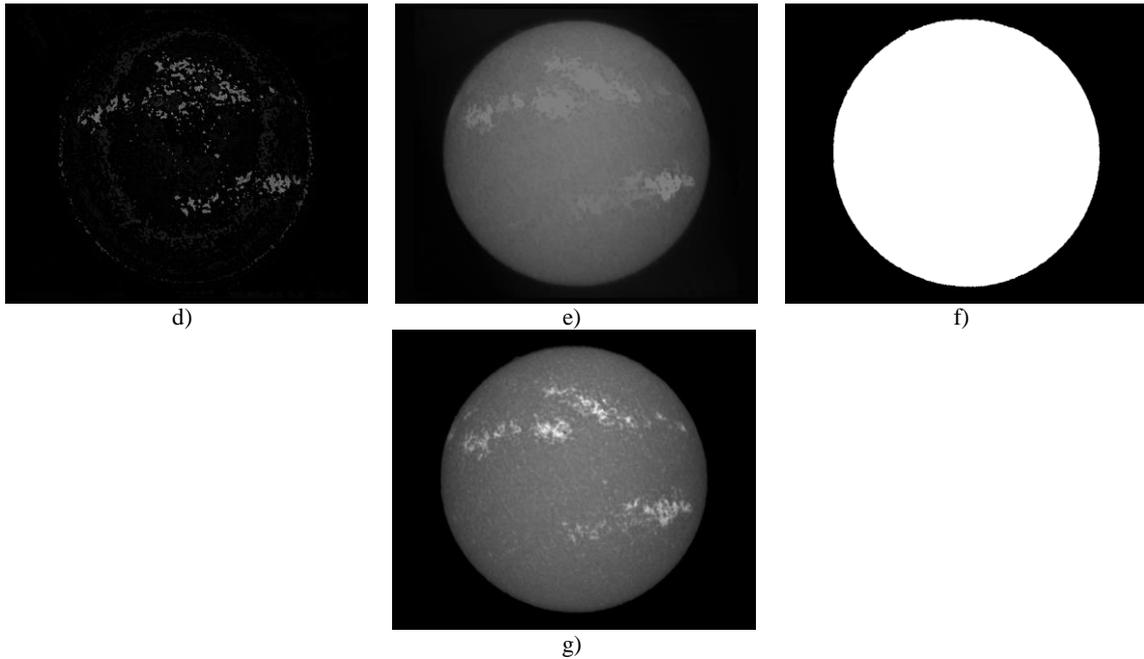

Figure 2 – Identification of solar disk in spectroheliograms: a) original image of a spectroheliogram; b) area closing of (a); c) closing of (b); d) difference between (a) and (c); e) reconstruction operation; f) binarization of (e); g) final image, obtained by the multiplication of the original image (a) and (f).

As illustrated in Figure 1 at the edge of the solar disk the brightness decreases sharply (Figure 3 (a)). For this reason, there are not many pixels with brightness similar to those on the edge. This means than on a pixel intensity histogram (Figure 3 (b)) the brightness of transition region falls in a local minimum, located with the 80$^{th}$ percentile of the histogram. This value is used to perform the threshold of the original image. The resulting binary image is shown in Figure 4(a).

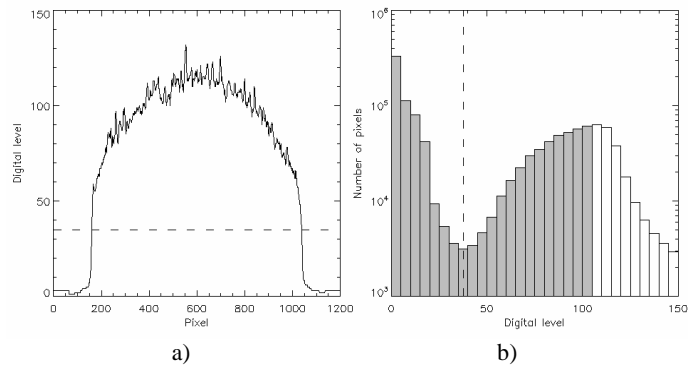

Figure 3 – Digital levels: a) Digital level of the central horizontal line of the image of Figure 1, with the minimum level of the histogram (dashed line); b) histogram of the digital levels of the image of Figure 1 (dashed line represents the minimum digital level within the 80$^{th}$ percentile);

The binary image (Figure 4 (a)), is used to determine the solar disk center coordinates (Xc, Yc) and its radius. Figure 4 shows the solar disk and the respective contour of the spectroheliogram of Figure 1.

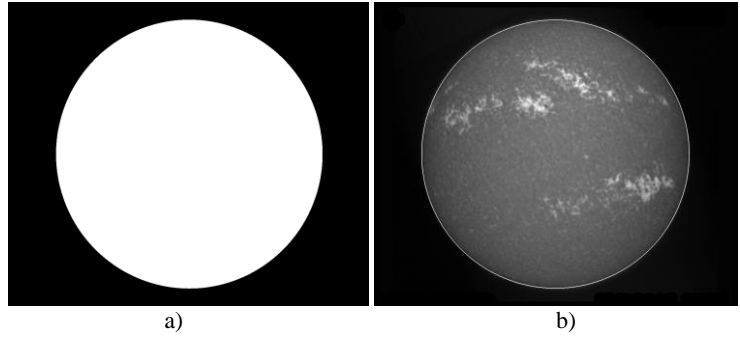

a)  b)

Figure 4 – Solar disk region and respective contour: a) binary image of the spectroheliogram of Figure 1 and b) spectroheliogram with the superimposed of the solar disk contour.

The coordinates of the center of the solar disk ($Xc, Yc$) were computed using the standard procedure to compute the centroid of an object:

$$Xc = \frac{\sum_{i,j} j*TH_{i,j}}{\sum_{i,j} TH_{i,j}} \quad (9)$$

and

$$Yc = \frac{\sum_{i,j} i*TH_{i,j}}{\sum_{i,j} TH_{i,j}} \quad (10)$$

Where $i$ and $j$ are, respectively, the number of rows and columns of the image, while T$H_{i,j}$ is the digital level of the threshold image at the pixel located at the $(i,j)$. The solar disk radius ($R$) is computed assuming that the disk is a perfect circle:

$$R = \sqrt{\frac{\sum_{i,j} TH_{i,j}}{\pi}} \quad (11)$$

A comparison of the solar radius computation for all spectroheliograms of solar cycle 24 (with radius around 450 pixels) show a great consistency of the results. Figure 5 shows the average difference between the estimated radius and the one to be expected from a sinusoidal fit is about one pixel, with a Root Mean Square (RMS) of 6.14 RMS. This radius difference is significantly smaller than the size of the solar disk in the images evaluated here. Notice that for instance a maximum 2 pixels uncertainty on the characterization of the solar disk's radius and central position generally results on a determination of the area of facular regions with a relative uncertainty around 5%. The determination of the center and the radius will be used to identify the facular regions in the hemispheres (north and south).

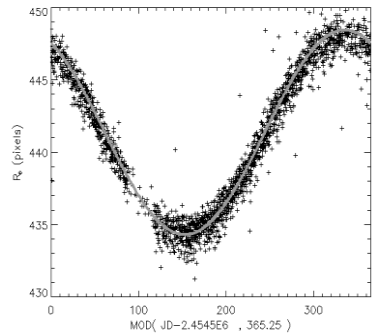

Figure 5 - Disk radius for 2083 spectroheliograms taken at the Coimbra observatory during solar cycle 24.

### 3.3. Morphological detection of facular regions

The principal aim of the algorithm to detect facular regions is to be completely automatic, i.e. to operate without any human intervention. The algorithm starts by applying a white top–hat by a disk of size 50 to the clean image, i.e. the image with a homogeneous background and without letters (image of Figure 2 (g). The objective of this operation, as observed in Figure 6 (a), is to have an image with only the areas that are brighter than their surroundings. After the top–hat operation, a hole filling was performed to connected and filled the areas inside facular regions. The results of this operation can be observed in Figure 6 (b). At this point, all facular regions are identified but it is necessary to isolate them, since not all bright areas correspond to facular regions; there are small or isolated bright pixels in the image that remains. This can be done through the application of a threshold (Figure 6 (c)). To have a completely automatic method several threshold values were tested according to the solar activity during a solar cycle and were integrated in the algorithm. For the years of maximum activity, the value used is the digital level of 30, to include sunspots, characterized by lower digital levels, since they correspond to dark pixels on the solar disk. For the years of the minimum solar activity the threshold value used is the digital level 50, due the fact that facular regions are less frequent and almost spotless. The digital level of 35 was used for the intermediate years of solar activity. The maximum value for all mentioned situations was fixed at 180, since the maximum digital level for the spectroheliograms rarely reach this value (see Figure 3 (a)).

The next step of the algorithm is to remove small areas (isolated pixels) that resisted to the threshold operation but do not correspond to facular regions. This can be done through the application of an erosion of the threshold image by a disk of size 2, followed by a reconstruction. Figure 6 (d) shows the result of the erosion operation. Figure 6 (e) shows the reconstruction operation, which preserve the bright areas that correspond to facular regions and simultaneously remove small objects. This image will be the image that contains the markers of facular regions, (Figure 6 (g)) after multiplying by the image that contains the contour of the solar disk (Figure 6 (f)). This ensures that all regions that will be detected by the segmentation are contained in the solar disk region. In this algorithm the segmentation used is the watershed transform applied to the image. Figure 6 (h) shows the catchment basins of the watershed operation applied to the image the clean image, this is, the image of Figure 2 (g)) e.g. using the image of the markers (Figure 6 (g) to impose the image regions to be segmented. Usually, when the watershed transform is applied directly to an image produce an "over–segmented" result. This is a well-known phenomenon which causes an excessively segmented image, that can be avoid by filtering previously the input image or, alternatively, using markers of the structures to be extracted. This last option was used in this segmentation algorithm. The gradient operation applied to the image of the watershed basins, allows to obtain the contours of facular regions (Figure 6 (i)). The result of the algorithm of automatic detection of facular regions can be observed in the image of Figure 7. This image shows the contours of facular region superimposed to the original image.

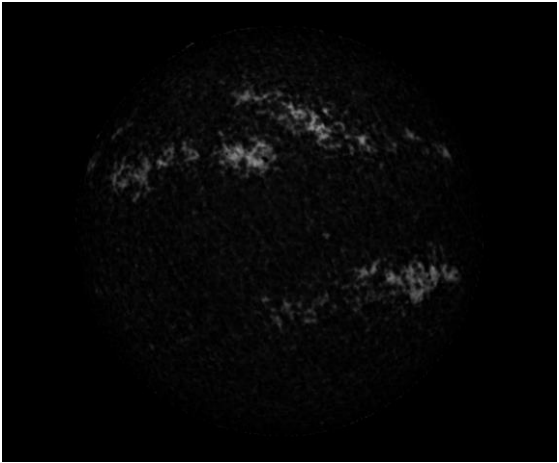

a) White top–hat applied to the image of Figure 2 (g).

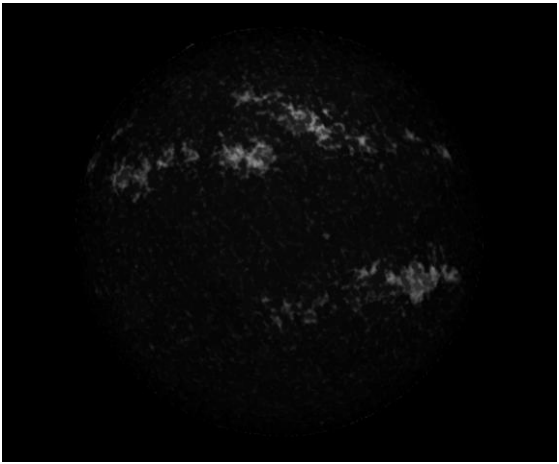

b) Hole fill operation.

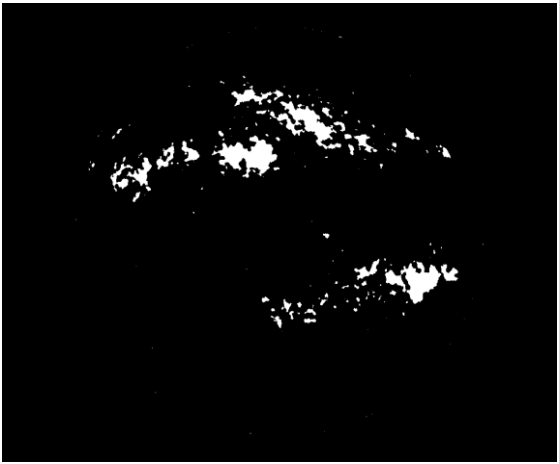

c) Threshold image.

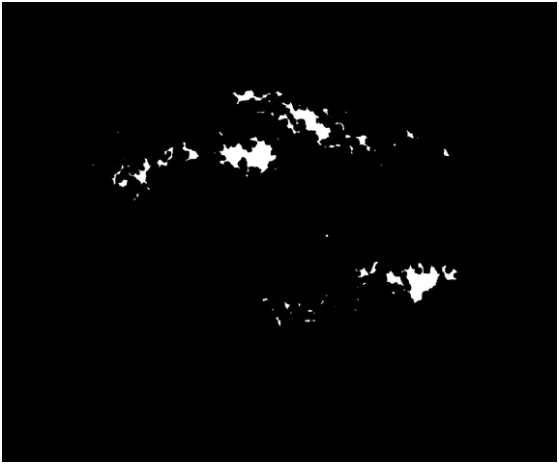

d) Erosion of the threshold image.

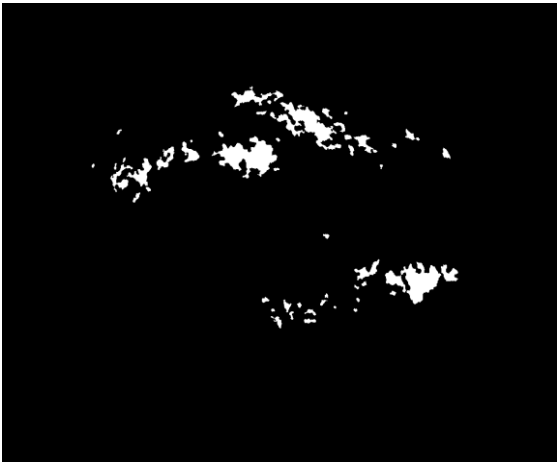

e) Reconstruction.

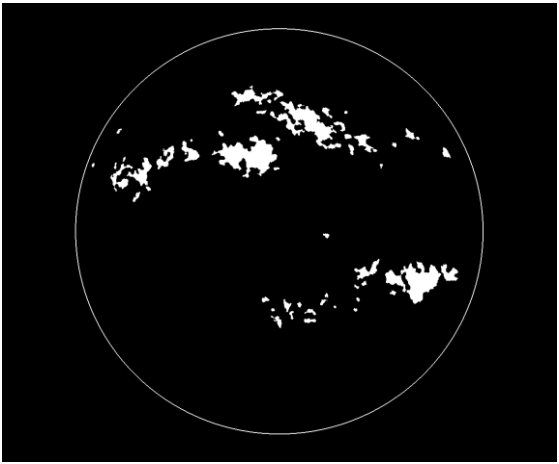

f) Reconstructed image with the contour of solar disk.

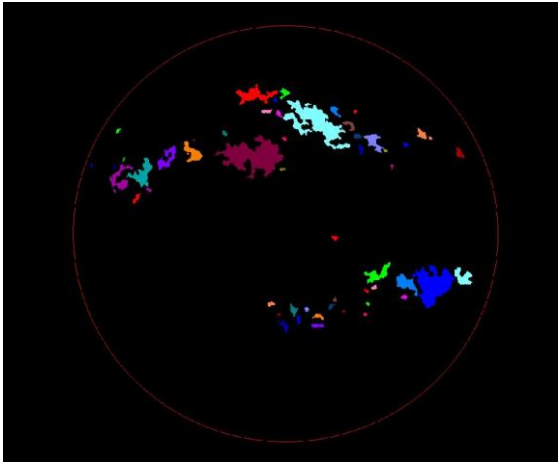

g) Markers of the facular regions.

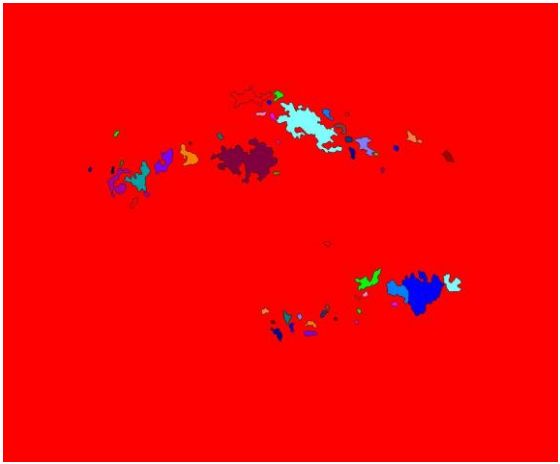

h) Basins obtained by the watershed operation, that corresponds to facular regions.

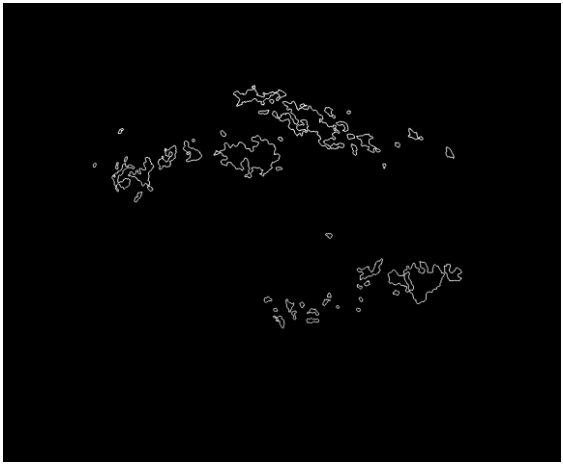

i) Contours of facular regions.

Figure 6 – Algorithm of automatic detection of facular regions.

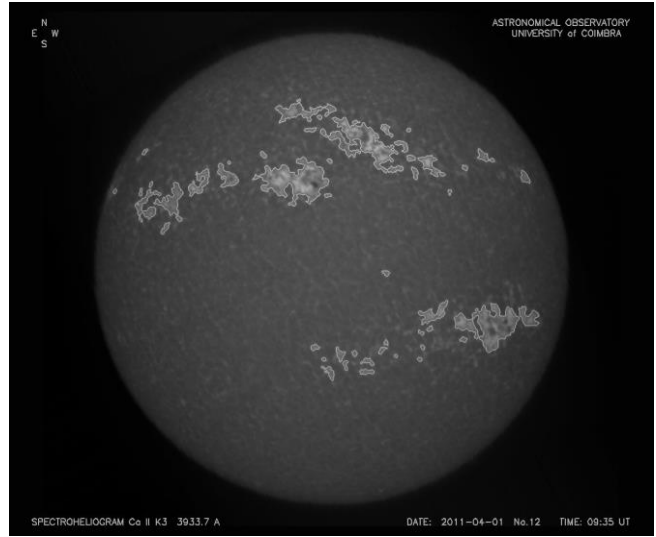

Figure 7 – Facular regions superimposed of the original image.

**3.4. Identification of the Solar hemispheres and determination of the area of facular regions**

After the identification of the facular regions the algorithm continues to determinate the facular regions in the northern and southern solar hemisphere, which can be useful to determinate areas and/or to determinate north and south asymmetries (Gonçalves et al., 2014).

Each pixel within the solar disk can be associated a set of heliocentric Cartesian coordinates *x* (number of pixels to the right of the center of the solar disk) and *y* (number of pixels above the center of the solar disk), as seen in Thompson (2006). These are the pixel's coordinates in respect to the center of the solar disk. It is possible to add a third coordinate *Z* perpendicular to the other two, through the expression:

$$Z^2 = X^2 \times Y^2 \qquad (12)$$

This Cartesian system can be converted to a spherical system of coordinates with latitude ($\theta$) and longitude ($\phi$):

$$\theta = arctan(Z/\sqrt{X^2+Y^2}) \qquad (13)$$

$$\emptyset = arg(X,Y) \qquad (14)$$

in which the *arg(X,Y)* function solves the quadrant ambiguity of the $\tan^{-1}$(*Y/X*) function (Thompson, 2006).

Nowadays Coimbra's spectroheliograms are rotated so the north/south poles appear at the top/bottom of the image. However, before 2002 the Sun was an inclination (P*) of the Sun's rotation axis in respect to the North–South line (North G.). In those cases, the coordinate system needs to be rotated along the Z axis to a new cartesian system (x', y' and z'), so at the pole one gets x' = 0.

On the other hand, due to the ecliptic's inclination in respect to the solar equator the Earth is generally above or below the solar equator. In these cases, the center of the solar disk is not at the equator, but at a latitude *Bo*. This heliographic latitude of the center of the solar disk is a solar ephemeris that can be easily computed (e.g Meeus, 1998). A *-Bo* rotation along the X axis produces a new cartesian coordinate system (x", y" and z"), in which y" = 0 for all points located at the solar equator. This new coordinate system was used to identify the solar northern and southern solar hemispheres (selecting the pixels for which y" > 0 or y" < 0).

The determination of each image's area of facular regions is done by integrating the image that contains the identified regions. Notice however, that towards the solar limb, the area of solar disk covered by each pixel increases by a factor proportional to 1/cos ($\theta d$), in which $\theta d$ is the angular distance between the pixel of coordinates (*X,Y*) and the center of the solar disk:

$$\theta_d(X,Y) = \sin\left(\sqrt{(X^2+Y^2)}/R\right) \quad (15)$$

thus, the total area of facular regions is expressed as a fraction of the visible solar surface (i.e. $2\pi R^2$, in which $R$ is the radius of the solar disk).

A similar approach can be used to compute the area of facular regions at the northern/southern hemispheres. In this case the image containing the facular regions identified should be multiplied by a mask containing the northern/southern hemispheres. Figure 8 shows the facular regions of the spectroheliogram of Figure 2, in the north and south hemisphere.

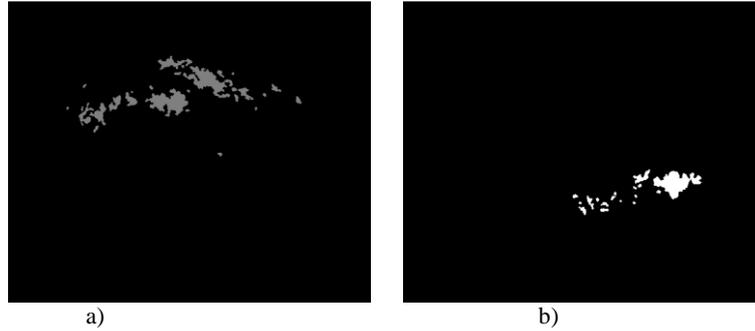

a) b)

Figure 8 – Facular regions on 1st April of 2011: a) North hemisphere and b) South hemisphere.

## 4. DATA ANALYSIS AND DISCUSSION

The automatic algorithm to detect facular regions was applied to the Coimbra spectroheliograms of the solar cycles 24 and 23. While the application of the algorithm for cycle 24 had the purpose of automatic detection of facular regions, for cycle 23 the objective was to obtain results that allow to validate the method. These images were compared with the images of cycle 23 already processed by the method developed by Dorotovič et al. (2007, 2010).

### 4.1. Illustration of the results and robustness:

Some results of the application of the algorithm for the cycle 24 are shown in the Figure 9.

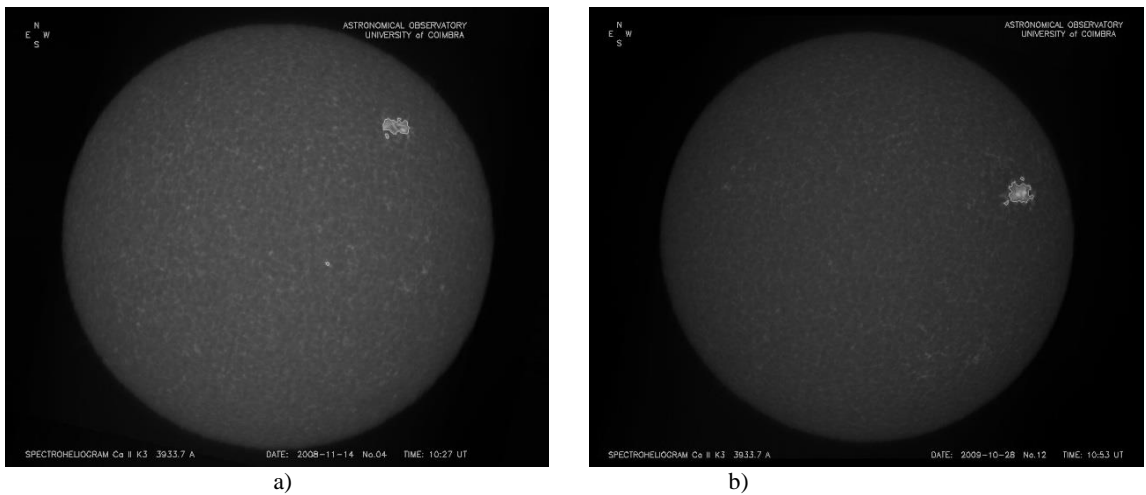

a) b)

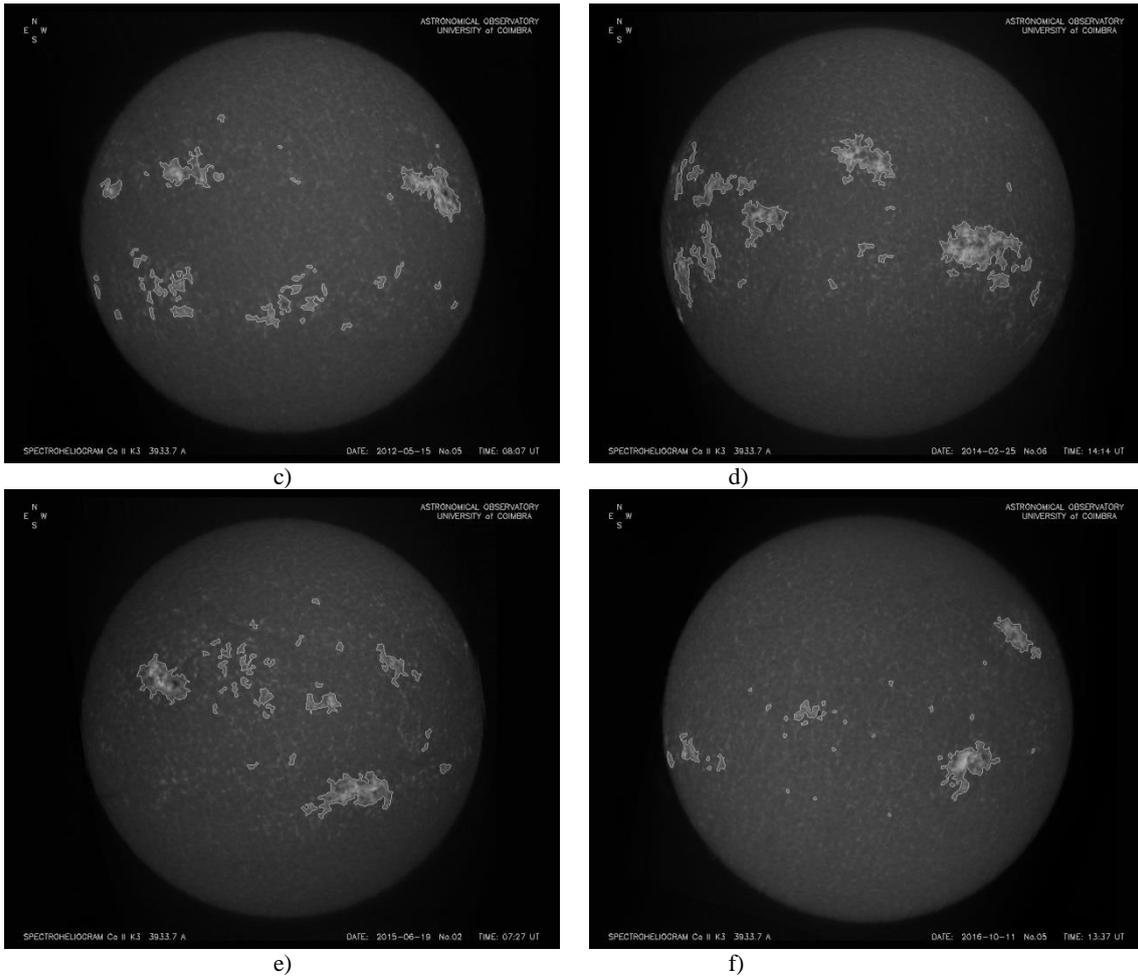

c)                                                        d)

e)                                                       f)

Figure 9 – Automatic detection of facular regions applied to spectroheliograms acquired during cycle 24: a) 14 November 2008; b) 28 October 2009; c) 15 May of 2012; d) 25 February of 2014; e) 19 July of 2015 and f) 11 October of 2016.

The application of automatic methods to ground-based images present some specific difficulties, due the Earth's atmosphere and meteorological factors. The good performance of the algorithm maintains when applying to images with strong atmospheric effects, as can be shown in the images of Figure 10.

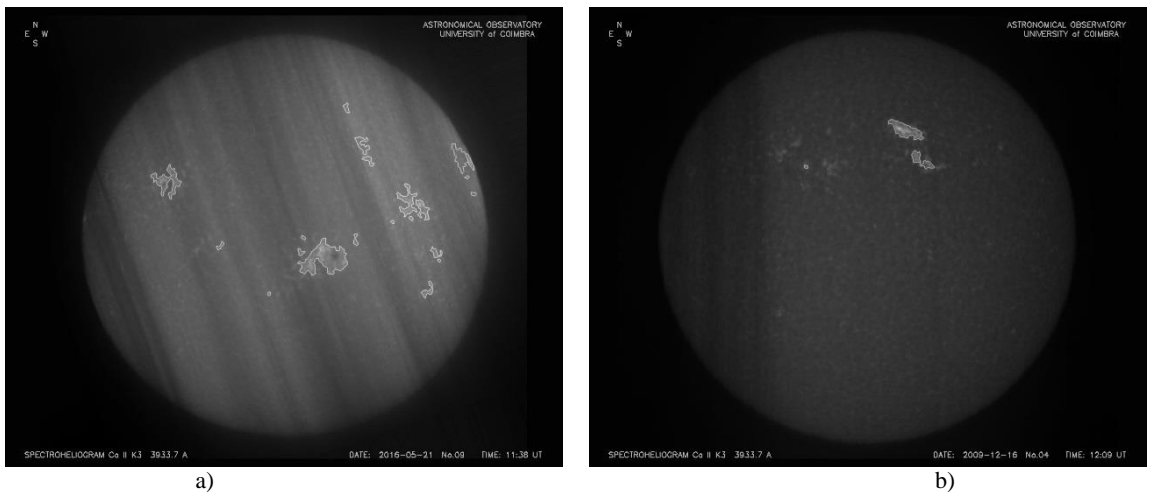

a)                                                       b)

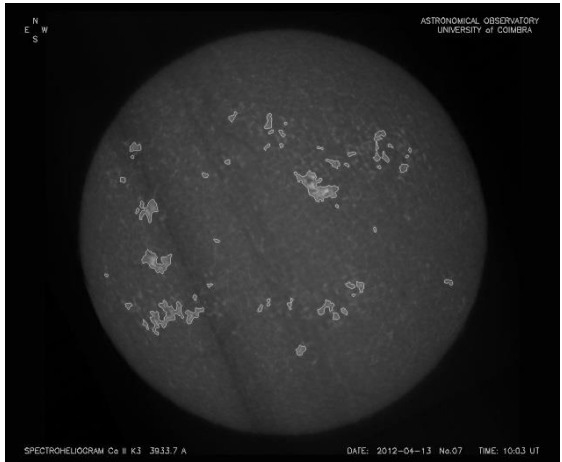
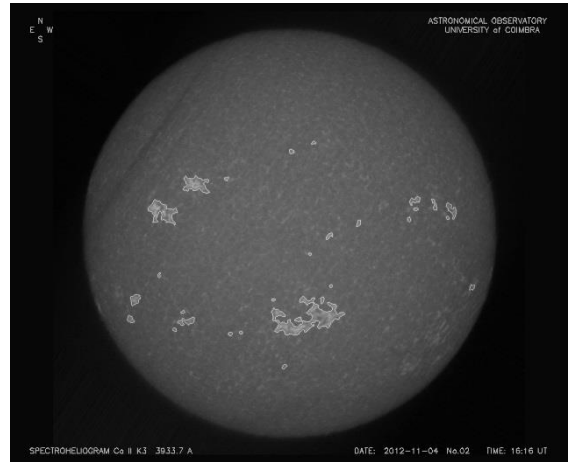
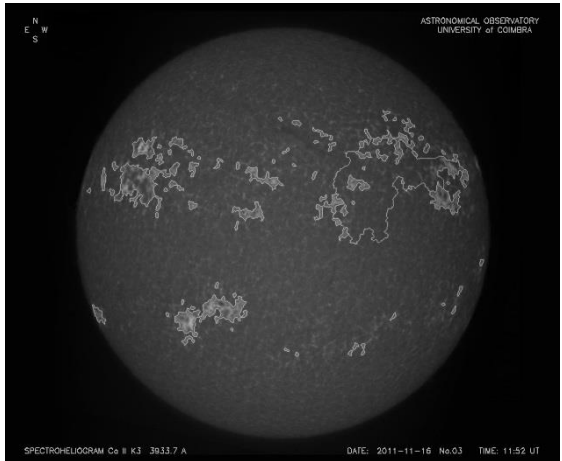
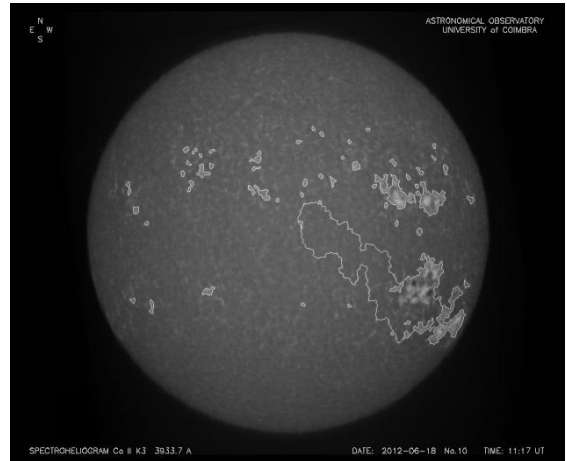

c)                  d)

Figure 10 – Automatic identification of facular regions in images with strong atmospheric effects: a) original image acquired at 2 May 2016; b) image acquired 16 December 2009; c) image acquired at 13 April 2012; d) image acquired 4 November 2012.

For cycle 24 the algorithm was applied approximately to 2083 spectroheliograms, from 2008 to 2016. The results were evaluated by an experienced astronomer that has estimated an error of bad identification of facular regions on about 7% of images. However, given the large amount of spectroheliograms that were analysed, it can be considered that the algorithm allows to obtain excellent results. Representative examples of wrong facular regions detection can be observed in the images of Figure 11.

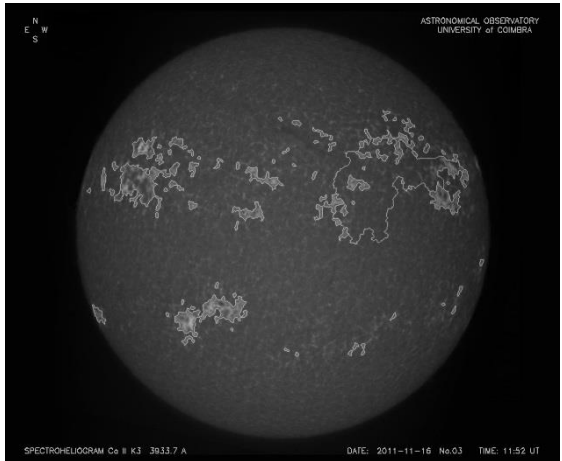
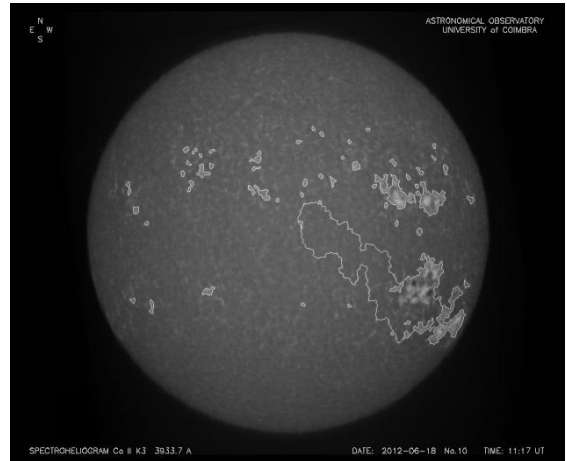

a)                  b)

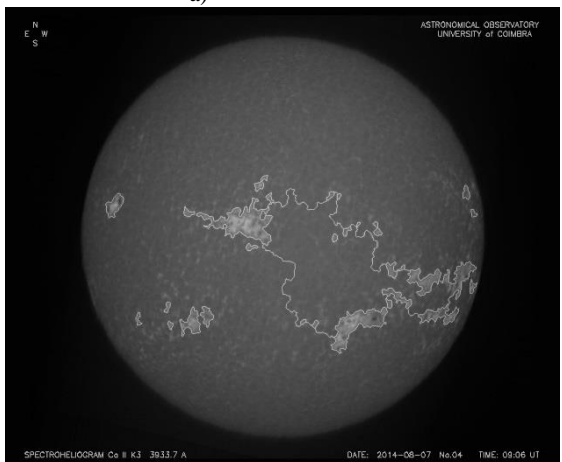
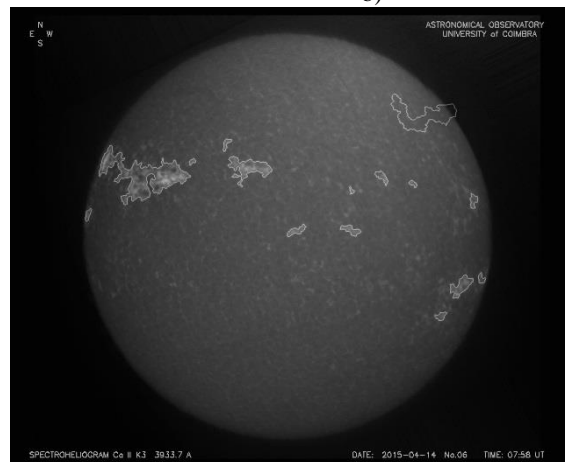

c)                  d)

Figure 11 – Erroneous results of automatic identification of facular regions: a) original image acquired at 16 November 2011; b) image acquired 18 June 2013; c) image acquired at 7 August 2014; d) image acquired 14 April 2015.

The typical errors are due to false segmentation of facular regions, on the application of the watershed transform. The marker image can contain an isolated pixel (not removed in previous steps) which lead to an erroneous segmentation. The type of error encountered was always caused by excessive segmentation (by identifying regions faculares more than real), and a reduce number of very small facular regions not always identified. The analysis of images acquired on consecutive days is a good method to evaluate the results. Figure 12 shows the results obtained in the days before and after of the spectroheliogram of Figure 11.

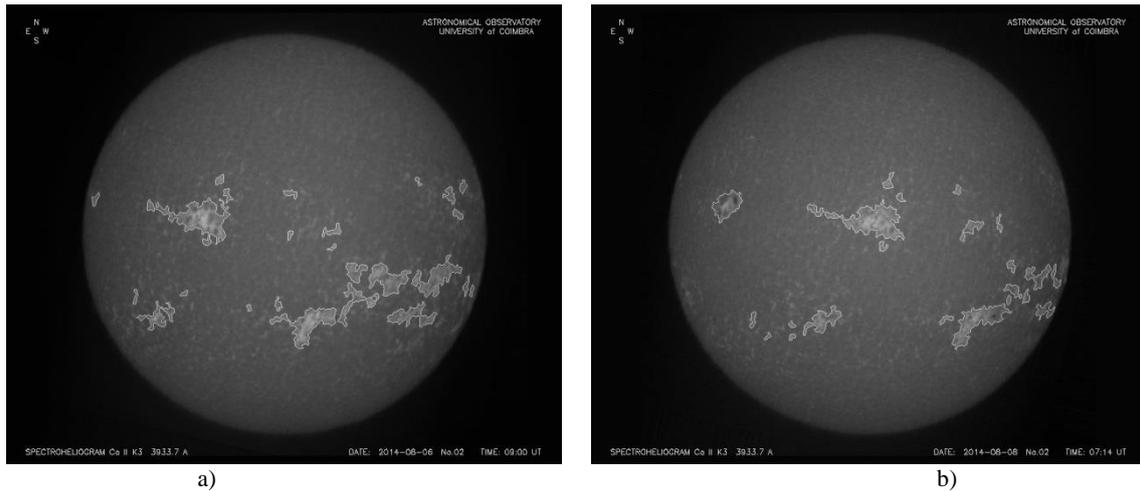

a)                  b)

Figure 12 – Results of automatic identification of facular regions: a) original image acquired at 6 August 2014; b) image acquired 8 August 2014.

In respect to the existence of outliers resulting from our approach, an evaluation of facular areas computed for this period showed only five clear outliers, as shown in Figure 13.

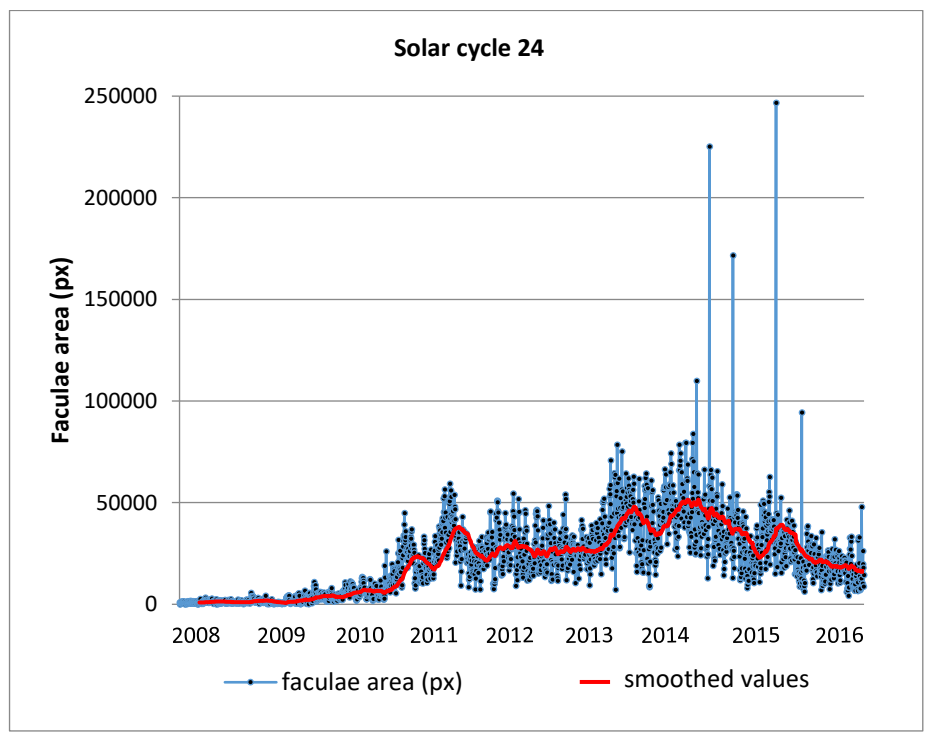

Figure 13 – Daily facular area during Solar cycle 24.

One possible interpretation of these outliers is due to problems within the images themselves (low signal–to–noise observations) or false segmentation (e.g. Figure 12). Given the total number of observations involved, one can consider this amount negligible.

**4.2. Validation on cycle 23**

Dorotovič et al. (2007, 2010) have developed a software tool to determine automatically the North-South Asymmetry Index of the Area of Ca II K Emission FEATures (AKFEAT), of the Area of Bright CHromospheric Features (ABCHF), from the OGAUC spectroheliograms. The algorithm is limited to determining the area of bright features in the emission line of CaII K3, in both hemispheres separately, after defining a threshold value of the relative brightness. The software starts with the determination of the quiet–Sun center–to–limb variation (CLV) using a method presented by Brandt and Steinegger (1998). The algorithm goes further, calculating the ABCHF as a sum of the area above the corresponding local quiet–Sun intensity. Finally, the algorithm also estimates the real area of bright chromospheric features, i.e. it transforms the area in fraction of the solar hemisphere (Figure 14). Further details on the method can be found in the cited papers.

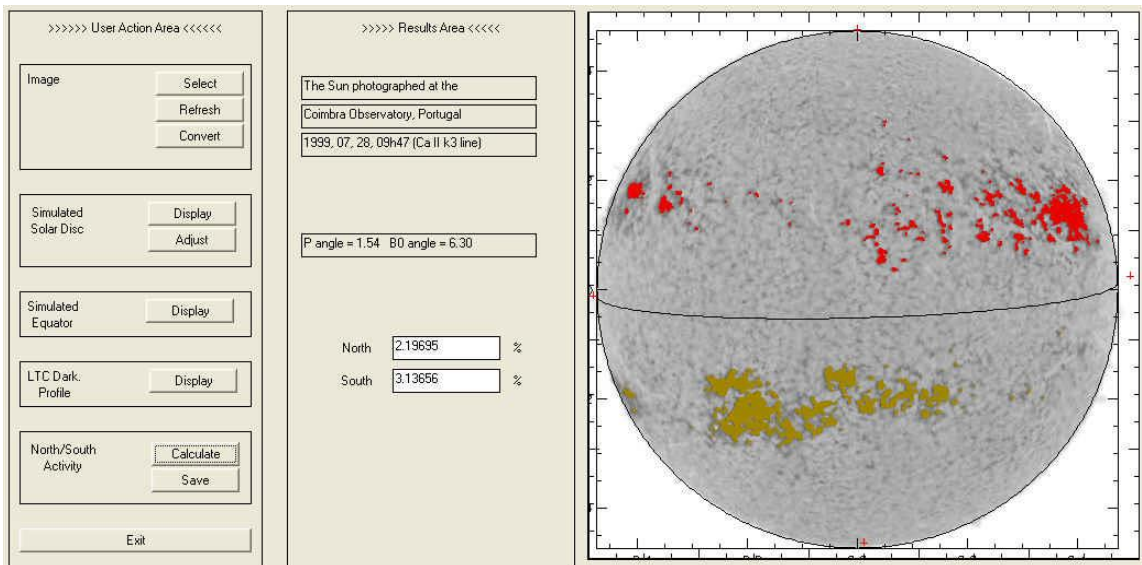

Figure 14 – Interface of the AKFEAT tool developed by Dorotovič et al. (2007).

The results obtained by our algorithm were compared with those obtained by Dorotovič et. al. (2007, 2010). Images taken in January 1997, February 1998, March 1999, April 2000, May 2001 and June 2002 (a total of 134 images), were used to make this comparison. Some examples are presented in figure 15. It must have pointed out that the images for cycle 23 were acquired in photographic film and digitalized latter. This explains the difference in contrast in the images of cycle 23 and 24.

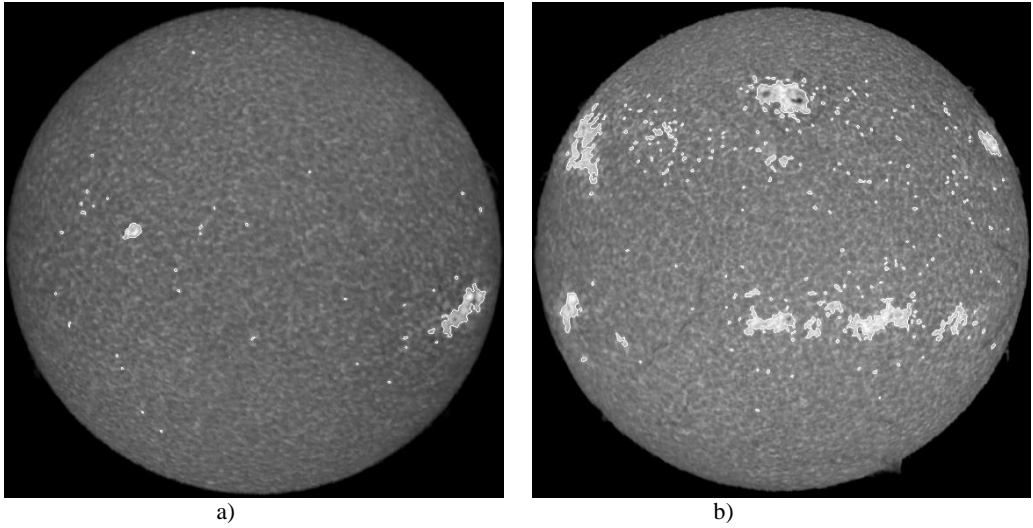

| a) | b) |

Figure 15 – Automatic detection of facular regions applied to spectroheliograms acquired during cycle 23: a) 7 August 1996 and b) 21 May 1999.

Figure 16 shows a comparison of the North, South and total areas of facular regions detected by both methods. As can be observed, the facular area estimates by the morphological method are well correlated with the values computed using the AKFEAT algorithm, with the Pearson's correlation coefficients above 0.8 (i.e. more than enough to have negligible false alarm probabilities). A linear regression to the area of southern facular regions (AreaS) computed using both methods yields:

$$AreaS_{AKFEAT} = 1.03 AreaS_{mat.morph.} - 0.19 \quad (16)$$

with an RMS = 0.61. This RMS is due to the larger dispersion of facular regions areas above 2% of the solar visible surface. One possible reason is due the different approach methods, particularly, in what concerns the use of thresholds. For the morphological method the thresholds values are fixed, while in the AKFEAT method, they vary from image to image. The AKFEAT method determines the reference (threshold) quiet-Sun intensity in a central circle of the solar disk, and in a set of concentric rings around it using a method of Brandt and Steinegger (1998). This yields different threshold values for each individual image. It is the main reason that the total area of facular regions derived using our method is slightly higher than the area of facular regions estimated using the AKFEAT method.

The mathematic morphology approach tended to identify slightly more northern facular regions (AreaN) than AKFEAT. Indeed, the best linear fit to both estimates is

$$AreaN_{AKFEAT} = 0.76 AreaN_{mat.morph.} + 0.03 \quad (17)$$

with an RMS = 0.49. Consequently, the total area of facular regions estimated using our method tends to be slightly higher than the one estimated using the AKFEAT software:

$$AreaT_{AKFEAT} = 0.84 AreaN_{mat.morph.} + 0.02 \, (RMS = 0,95) \quad (18)$$

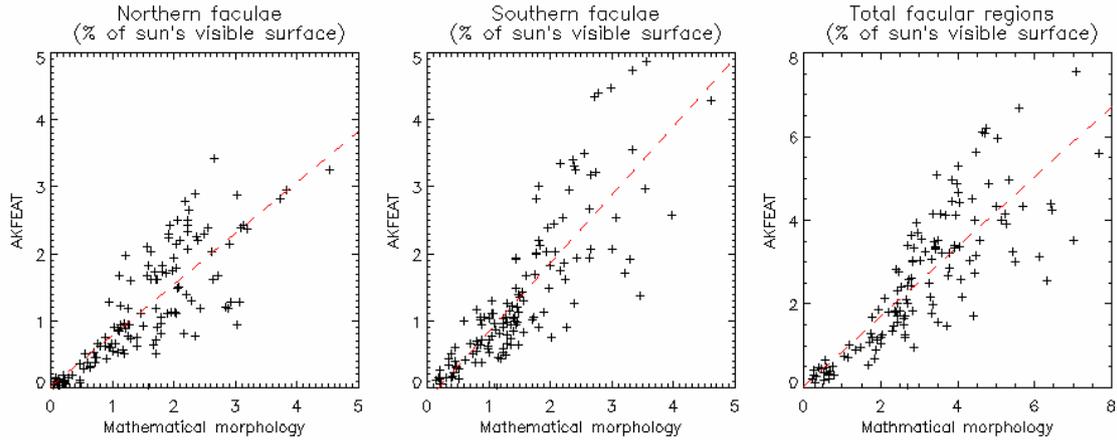

Figure 16 – Comparison between the fraction of the solar surface of facular regions during 6 months of solar cycle 23, identified using our method against the estimates of Dorotovič et al. (2007, 2010) software (AKFEAT). From left to right: area of the northern faculae, area of the southern faculae and total area of facular regions. The dashed lines correspond to the best linear fits to the data.

Considering that the AKFEAT software was designed to compute the N–S asymmetry of facular regions, a comparison between both algorithms was performed (based on the difference facular areas of the North and South hemisphere) (Figure 17). Again, there is a significant degree of correlation between both estimates (r = 0.83), for which the best linear fit corresponds to:

$$(AreaN - AreaS)_{AKFEAT} = 1.08(AreaN - AreaS)_{mat.morph.} - 0.19 \quad (RMS = 0.60) \quad (19)$$

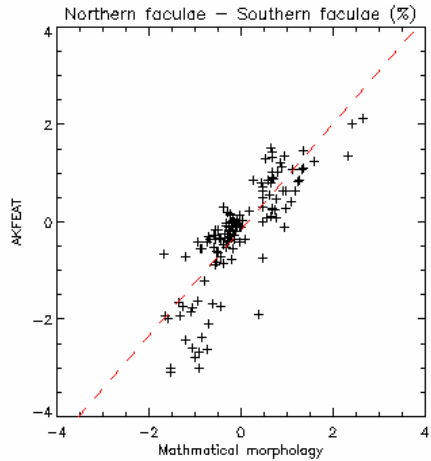

Figure 17– Comparison between the North–South asymmetry of facular regions computed using our methodology and the AKFEAT algorithm. The dashed line corresponds to the best linear fit to the data.

**4.3. Comparison with ASSA method (SDO HMI images)**

The Automatic Solar Synoptic Analyzer (ASSA) method is a free available software developed by the Korean Space Weather Center of the Radio Research Agency. The results obtained by the ASSA method were qualitatively compared with the results obtained by the application of the automatic method to the HMI SDO images (section 3.4), by an expert observer. This qualitative validation was performed for a representative set of images. Figure 2018 shows examples of the comparison of the results obtained by both methods. Some examples of application of the algorithm are shown in Figure 18.

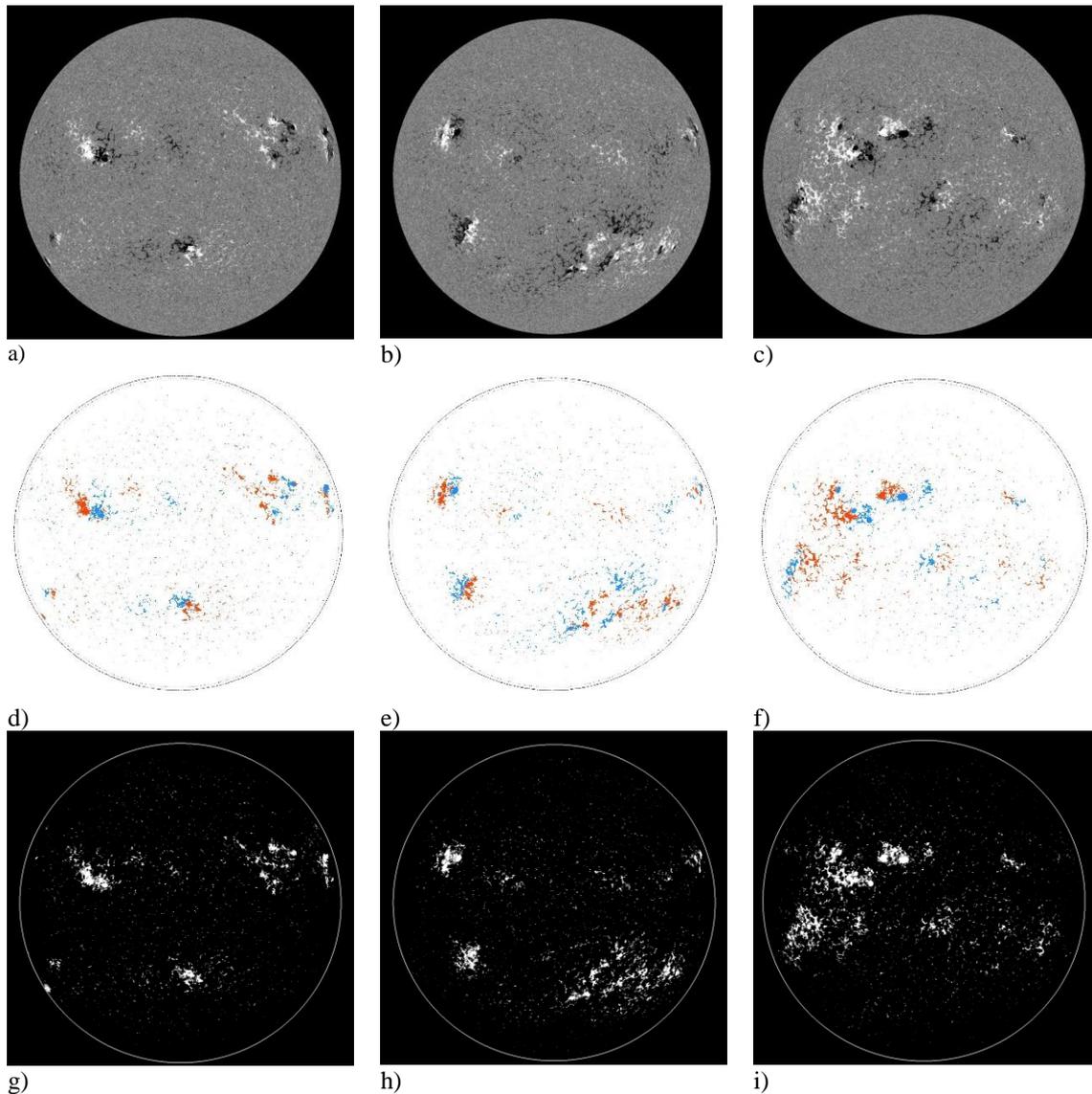

Figure 18 – HMI SDO images showing the polarity of magnetic field: a) original image acquired at 3 September 2011; b) original image acquired 10 August 2012; c) original image acquired at 10 May 2014; d) ASSA results of the image (a); e) ASSA results of the image (b); f) ASSA results of the image (c); g) automatic detection of facular regions of the image (a); h) automatic detection of facular regions of the image (b); i) automatic detection of facular regions of the image (c).

As can be observed from figure 18, the results obtained using both methods are in conformity in what concerns the areas detected. The only difference is that the morphological method does not determinate the magnetic polarity of the active regions, due the lack of magnetic information in Coimbra's data.

### 4.4. Application to Kharkiv spectroheliograms

The algorithm was applied to spectroheliograms from the Astronomical Observatory of Kharkiv State University, available through the database http://www.astron.kharkov.ua/ssm/. The objective was to test the applicability of the method to other images, using the same parameters for morphological operators and threshold values. Figure 19 shows an example of the results obtained for the Ca KII Kharkiv spectroheliograms.

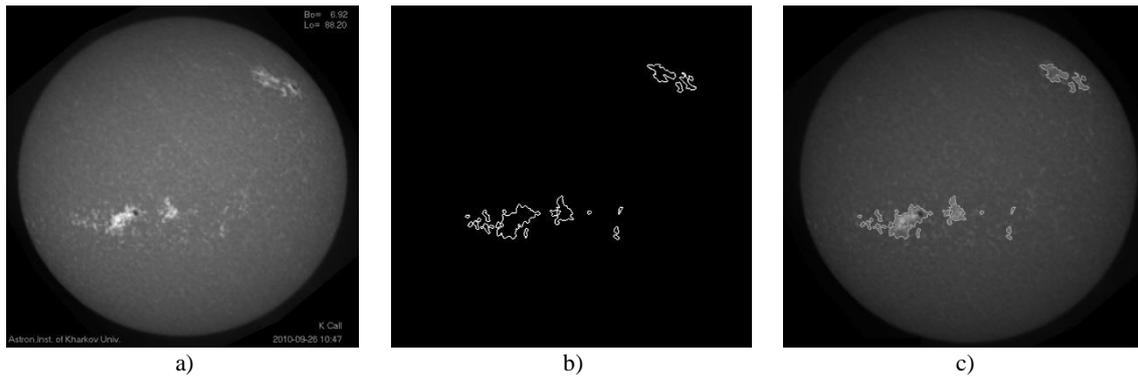

| a) | b) | c) |

Figure 19 – Automatic detection of facular regions applied to the Kharkiv spectroheliogram acquired at 26 September 2010: a) original image; b) facular regions obtained by the automatic method and c) facular regions superimposed to the original image.

**5. CONCLUSIONS**

One of the major challenges in the development of automatic methods is their universal applicability. In other words, to have procedure that can be applied to different of types images (high resolution, low resolution) and from different sources (space missions, ground observations) with the same purpose. In this case the method described here was applied to spectroheliograms acquired from different observatories and SDO HMI images, promising results.

A mathematical morphology algorithm was developed to be applied to the CaII K3 series spectroheliograms of OAGUC, with the purpose of creating an automatic method to detect the chromospheric plages during the solar cycle 24. The tool was quantitatively compared with the results for cycle 23 performed by Dorotovic et al. (2007, 2010), and qualitatively with ASSA model applied to SDO HMI magnetograms.

In what concerns to ground–based images, the application of automatic methods can present some additional difficulties, due the Earth's atmosphere and meteorological factors, occasionally providing an additional source of noise to our data or adding artifacts not present in the Sun and the limb darkening correction. It is important to highlight that the algorithm developed in this work does not need to pre–process the images to remove the atmospheric effects, the intensity and contrast. Additionally, the procedure proposed here has the added advantage of being unaffected by the limb darkening effect and thus requiring no pre–processing of the data with the sole purpose of removing it. Moreover, the results obtained from the morphological transforms agree with the results obtained from other approaches, including on images obtained with atmospherics artifacts (e.g. some clouds).

The versatility of the automatic methods is high, not only due to its applicability to any type of image (from high resolution to spectroheliograms, for example), but also for its extraction power of characterizing and quantifying parameters of the solar activity.

As future work it is intended to analyze the whole Coimbra's data series and make the entire data set public. It is expected that this analysis will be a valuable contribution to study of the variability of the Sun's activity, namely the North-South asymmetry and the correlation between solar and geomagnetic activities.

**Acknowledgements:**
This research tasks are performed in the frame of a mobility project of SRDA (APVV) SK–PT–2015–0004 supporting cooperation between organizations in the Slovak Republic and Portugal (2016 – 2017). This study was also partially supported by the Portuguese Government through the Foundation for Science and Technology – FCT, CITEUC Funds (project: UID/Multi/00611/2013), CMUC Funds (UID/MAT/00324/2013) and FEDER – European Regional Development Fund through COMPETE 2020 – Operational Programme Competitiveness and Internationalization.
Sara Carvalho's work has been funded by FCT grant SFRH/BD/107894/2015. Ana Lourenço´s work was supported by the project ReNATURE– Valuation of Endogenous Natural Resources in the Central Region (CENTRO–01–0145–FEDER–000007).
We are also very grateful for the constructive comments and suggestions provided by the anonymous reviewers.

# REFERENCES

Aboudarham, J., Scholl, I., Fuller, N., Fouesneau, M., Galametz, M., Gonon, F., Maire, A., Lero, Y., 2008. Automatic detection and tracking of filaments for a solar feature database. Ann. Geoph. 26, 243–248.

Aschwanden, M., 2010. Image Processing Techniques and Feature Recognition in Solar Physics, Solar Phys. 262, 235–275.

Ayres, T., Longcope, T., 2012. Ground–based Solar Physics in the Era of Space Astronomy, Whitepapper, 2012 Heliophysics Decadal survey.

Barata, M. T., Lopes, F., Pina, P., Alves, E. I., Saraiva, J., 2015. Automatic detection of wrinkle ridges in Venus Magellan imagery. In: T. Platz, M. Massironi, P. K. Byrne and H. Hiesinger (eds), Volcanism and Tectonism Across the Inner Solar System, Geologycal Society of London, Special Publications, 401, 357–376, first published on January 9, 2014, doi:10.1144/SP401.5.

Benkhalil, A., Zharkova, V., Ipson, S., Zharkov, S., 2004. Automatic Detection of Active Regions on Solar Images. Lect. Not. Comp. Sci. 3215, 460–466.

Beucher, S., 1990. Segmentation d'images et morphologie mathématique. Thèse de Doctorat, ENSMP, Paris.

Borda, R., Mininni, P., Mandrini, C., Gómez, D., Bauer, O., Rovira, M., 2002. Automatic solar flare detection using neural network techniques. Solar Phys. 206, 347–357.

Brandt, P.N., Steinegger, M., 1998. On the Determination of the Quiet–Sun Center–to–Limb Variation in Ca II K Spectroheliograms. Solar Phys. 177, 287 –294.

Caballero, C., Aranda, M.C., 2014. Automatic Tracking of Active Regions and Detection of Solar Flares in Solar EUV Images. Solar Phys. 289, 1643–1661.

Carvalho, S., Pina, P., Barata, T., Gafeira, R., Garcia, A., 2015. Ground–based observations of sunspots from the Observatory of Coimbra: evaluation of different automated approaches to analyse its datasets, Coimbra Solar Physics Meeting: Ground–based Solar Observations in the Space Instrumentation Era, Dorotovic, I.; Fischer, C.; Temmer, M. (eds.), Astronomical Society of the Pacific Conference Series Vol. 504, 125–131.

Curto, J.J., Blanca, M., Martínez, E., 2008. Automatic sunspots detection on full-disk solar images using mathematical morphology. Solar Phys. 250, 411–429.

Denker, C., Johannesson, A., Marquette, W., Goode, P., Wang, H., Zirin, H., 1999. Synoptic H full–disk observations of the Sun from Big Bear Solar Observations. Solar Phys. 211, 83–102.

Dorotovič, I., Journoud, P., Rybák, J., Sýkora, J., 2007. NorthSouth Asymmetry of Ca II K Plages, in Proc. Coimbra Solar Physics Meeting on ' The Physics of Chromospheric Plasmas' Petr Heinzel, Ivan Dorotovic and Robert J. Rutten, eds., ASP Conference Series, Vol. 368, 527.

Dorotovič, I., Rybák, J., Garcia, A., Journoud, P., 2010. North–south asymmetry of Ca II K regions determined from OAUC spectroheliograms: 1996 – 2006, Proceedings of the 20th National Solar Physics Meeting, 31 May – 4 June, 2010 Papradno, Slovakia, 58–63.

Dorotovič, I., Shahamatnia, E., Lorenc, M., Rybanský, M., Ribeiro, R. A., Fonseca, J. M., 2014. Sunspots and coronal bright points tracking using a hybrid algorithm of PSO and active contour model. Sun and Geosph. 9 (1–2), 81–84.

Falconer, D., Barghouthy, A., khazanv, I., Moore, R., 2011. A tool for empirical forecasting of major flares, coronal mass ejections, and solar particle events from a proxy of active–region free magnetic energy. Space Weather, 9, S04003.

Fonte, C., Fernandes, J., 2009. Application of fuzzy sets to the determination of sunspot areas. Solar Phys. 260, 21–41.

Fuller, N., Aboudarhma, J., 2004. Automatic Tracking of solar filaments versus manual digitalization. LNCS, V. 3215, 467–475.

Fuller, N., Aboudarhma, J., Bentley, R., 2005. Filament recognition and image cleaning on Meudon H–alpha Spectroheliograms. Solar Phys., 227, 61–73.

Gafeira, R., Fonte, C. C., Pais, M. A., Fernandes, J., 2013. Temporal Evolution of Sunspot Areas and Estimation of Related Plasma Flows. Solar Phys. 289 (2), 1531–1542.

Garcia, A., Klvaňa, M., Sobotka, M., 2010. Measurements of chromospheric velocity fields by means of the Coimbra University spectroheliograph. Cent. Eur. Astr. Bull. 34, 47–56.

Gill, C. D., Fletcher, L., Marshall, S., 2010. Using active contours for semi–automated tracking of UV and EUV solar flare ribbons. Solar Phys. 262, 355–371.

Gonçalves, E., Mendes-Lopes, N., Dorotovic, I., Fernandes, J. M., Garcia, A., 2014. North and South Hemispheric Solar Activity for Cycles 21–23: Asymmetry and Conditional Volatility of Plage Region Areas. Solar Phys. 289 (6), 2283-2296.

Göker, U., Singh, J., Nutku, F., Priyal, M., 2016. A Statistical Analysis of Solar Surface Indices Through the Solar Activity Cycles 21–23, arXiv:1604.03011v2

Goussies, N., Mejail, M., Jacobo, J., Stenborg, G., 2010. Detection and tracking of coronal mass ejections based on supervised segmentation and level set. Pattern Rec. Lett. 31, 496–501.

Higgins, P., 2012. Sunspot group evolution and the Global Magnetic Field of the Sun, PhD Thesis, School of Physics, University of Dublin, Trinity College, 309pp.

Higgins, P., Gallagher, P., McAteer, R., Bloomfield, D., 2011. Solar magnetic feature detection and tracking for space weather monitoring. Adv. Space Res. 47, 2105–2117.

Hill, F., Burkepile, J., Choudhary, D.P., Giampapa, M., Keil, S., Goode, P., Kuhn, J., Leka, K.D., Pevtsov, A., Rhodes, E., Thompson, M., Ulrich, R., 2010. The need for synoptic optical solar observations from the ground, A Whitepaper submitted to the 2013–2022 Decadal Survey in Solar and Space Physics (Solar and Heliospheric Physics), ArXiv e–prints, 8pp.

Irbah, A., Bouzaria, M., Lakhal, L., Moussaoui, R., Borgnino, J., Laclare, F., Delmas, C., 1999. Feature Extraction from Solar Images Using Wavelet Transform: Image Cleaning for Applications to Solar Astrolabe Experiment. Solar Phys. 185, 255–273.

Kostik, R., Khomenko, E., 2014. Properties of convective motions in facular regions. Astr. & Astr. 10p. (https://arxiv.org/abs/1207.4340).

Lantuéjoul, Ch., Beucher, S., 1980. On the use of the geodesic metric in image analysis, Jour. Micr. 121, 29–49.

Nesme-Ribes E., Meunier, N., Collin, B., 1996. Fractal analysis of magnetic patterns from Meudon spectroheliograms. Astr. & Astr. 308, 213–218.

Manish, T.I., Murugan, D., Kumar, G. T., 2014. Automatic detection of sunspots activities using advanced detection model. IOSR Jour. Comp. Eng. 6 (2), 83–87.

Martens, P., Attill, G., Davey, A., Engell, A., Farid, S., Grigis, P., Kaper, J., Korreck, K., Saar, S., Savcheva, A., Su, Y., Testa, P., Willis–Davey, M., Bernasconi, P., Raouafi, N., Delouille, V., Cirtaisn, J., DeForest, C., Angryk, R., De Moortel, I., Wiegelmann, T., Georgoulis, M., McAteer, R., Timmons, R., 2012. Computer vision for the Solar Dynamics Observatory (SDO). Solar Phys. 275, 79–113.

Matheron, G., 1967. Éléments pour une théorie des milieus poreux. Masson, Paris, 168pp.

Matheron, G., 1975. Random sets and integral geometry. John Wiley & Sons, New York, 261 pp.

Meeus, J., 1998, Astronomical algorithms (2nd ed.) by J. Meeus. Richmond, VA, Willmann-Bell, 1998.

Meunier, N., Delfosse, X., 2009. On the correlation between Ca and Hα solar emission and consequences for stellar activity observations. Astr. & Astr. 501, 1103–1112.

Meyer, F., 1979. Cytologie quantitative et morphologie mathématique. Thèse de Doctorat, ENSMP, Paris.

Meyer, F., 1986. Automatic screening of cytological specimens. Comp. Vis. Graph. Im. Proc. 35, 356–369.

Olmedo, O., Zhang, J., Wechsler, H., Poland, A., Borne, K., 2008. Automatic detection and tracking of coronal mass ejections in coronagraph time series. Solar Phys. 248, 485–499.

Pérez–Suárez, D., Higgins, P., Bloomfield, S., McAteer, R. T., Krista, L., Byrne, J., Gallagher, T., 2011. Automated solar feature detection for space weather applications, in Applied Signal and Image Processing: Multidisciplinary Advancements (ed. Rami Qahwaji, Roger Green and Evor L. Hines), 207–225.

Qahwaji, R., Colak, T., 2005. Automatic detection and verification of solar features. Int. Jour. Im. Syst. Tech. 15, 199–210.

Qu, M., Shih, F., Jing, J., Denker, C., Wang, H., 2005. Automated detection and identification of solar filaments and sunspots. AGU Spring Meeting Abstracts, 6.

Scholl, I., Habbal, S., 2008. Automatic Tracking and classification of coronal holes and filaments based on EUV and magnetogram observations of the solar disk. Solar Phys. 248, 425–4.

Serra, J., 1982. Image analysis and mathematical morphology. Academic Press, London.

Soille, P., 2002. Morphological Image Analysis – Principles and Applications, 2nd Ed., Berlin, Springer–Verlag.

Solanki, S.K., Fligge, M., 1999. A reconstruction of total solar irradiance since 1700, Geoph. Res. Lett. 26 (16), 2465–2468.

Solanki, S. K., Unruh, Y.C. 2013. Solar Irradiance Variability Astron. Nachr. 334, 145–150.

Thompson, W. T., 2006. Coordinate systems for solar image data, Astr. & Astr. 449 (2), 791–803.

Verbeeck, C., Delouille, V., Mampaey, B., De Visscher, R., 2014. The SPoCA–suite: Software for extraction, characterization, and tracking of active regions and coronal holes on EUV images. Astr. & Astr. 561, A29.

Verbeeck, C., Higgins, P., Colak, T., Watson, F., Delouille, V., Mampaey, B., Qahwaji, R., 2013. A multi–wavelength analysis of active regions and sunspots by comparison of automatic detection algorithms. Solar Phys. 283, 67–95.

Veronig, A., Steinegger, M., Otrub, A., Hanslmeie, A., Messerotti, M., Temmer, M., Gonzi, S., Brunner, G., 2001. Automatic Image Processing in the Frame of a Solar Flare Alerting System. Hvar Obs. Bull. 24 (1), 195–205.

Walton, S. R., Chapman, G. A., Cookson, A. M., Dobias, J. J., Preminger, D. G., 1998. Processing Photometric Full–Disk Solar Images. Solar Phys. 179, 31–42.

Walton, S. R., Preminger, D. G., 1999. Restoration and Photometry of Full–Disk Solar Images. Astr. Jour. 514, 959–971.

Zhao, C., Lin, G., Deng, Y., Yang, X., 2016. Automatic recognition of sunspots in HSOS full–disk solar images, Publications of Astronomical Society of Australia (PASA), 9p.

Zharkov, S., Zharkova, V., Ipson, S., 2005. Statistical Properties of sunspots in 1996–2004: 1. Detection, North–South asymmetry and area distribution. Solar Phys. 228, 377–397.

Zharkova, V., Aboudarham, J., Zharkov, S., Ipson, S., Benkhalil, K., Fuller, N., 2005. Solar features catalogue in EGSO. Solar Phys. 228, 361–375.

Zharkova, V., Ipson, S., Benkhalil, A., Zharkov, S., 2004. Feature Recognition in solar images. Art. Int. Rev. 23, 209–266.